\documentclass[12pt,thmsa]{article}
\usepackage{sw20lart}

\input{tcilatex}
\begin{document}

\title{Form factors, branching ratio and forward-backward asymmetry in $%
B\rightarrow K_{1}\ell ^{+}\ell ^{-}$ decays}
\author{M. Ali Paracha, Ishtiaq Ahmed \\
Department of Physics and National Centre for Physics, Quaid-i-Azam\\
University, Islamabad, Pakistan. \and M.Jamil Aslam \\
COMSATS Institute of Information Technology, Islamabad, Pakistan.}
\maketitle

\begin{abstract}
We study long-distance effects in rare exclusive semileptonic decays $%
B\rightarrow K_{1}\ell ^{+}\ell ^{-}$ , $K_{1}$ is the axial vector
meson.The form factors, describing the meson transition amplitudes of the
effective Hamiltonian, are calculated using Ward identites which are then
used to calculate branching ratio and forward-backward asymmetry in these
decay modes. The zero of forward-backward asymmetry is of special interest
and provide us the precission test of Standard model.
\end{abstract}

\section{Introduction}

The investigation of rare semileptonic decays of the $B$ meson induced by
the flavor-changing-neutral-current (FCNC) transitions $b\rightarrow s$
provide potentially stringent tests of standard model (SM) in flavor
physics. In SM these FCNC\ transitions are not allowed at tree level but are
induced by the Glashow-Iliopoulos-Miani (GIM) amplitudes \cite{1} at the
loop level. Additionally these are also suppressed in SM due to their
dependence on the weak mixing angles of the quark-flavor rotation matrix $-$
the Cabibbo-Kobayashi-Maskawa (CKM) matrix \cite{2}. These two circumstances
make the FCNC decays relatively rare and hence important for the presence of
new physics, commonly known as physics beyond SM.

The experimental observation of inclusive \cite{3} and exclusive \cite{4}
decays, $B\rightarrow X_{s}\gamma $ and $B\rightarrow K^{*}\gamma $ , has
prompted a lot of theoretical interest on rare $B$ meson decays. However, in
case of exclusive decays any reliable extraction of the perturbative
(short-distance) effects encoded in the Wilson coefficients of the effective
Hamiltonian \cite{5,6,7,8,9} requires an accurate separation of the
nonperturbative (long-distance contributions), which therefore should be
known with high accuracy. The theoretical investigation of these
contributions encounters the problem of describing hadron structure, which
provides the main uncertainty in the predictions of exclusive rare decays.
In exclusive $B\rightarrow K,K^{*}$ decays the long-distance effects in the
meson transition amplitude of the effective Hamiltonian are encoded in the
meson transition form factors. Many exclusive $B\rightarrow K\left(
K^{*}\right) \ell ^{+}\ell ^{-}$\cite{10,11,12}, $B\rightarrow \gamma \ell
^{+}\ell ^{-}$\cite{13}, $B\rightarrow \ell ^{+}\ell ^{-}$\cite{14}
processes based on $b\rightarrow s\left( d\right) \ell ^{+}\ell ^{-}$ have
been studied in literature and many frameworks have been applied to the
description of meson transition form factors: among them the worth
mentioning are constituent quark models, QCD sum rules, lattice QCD,
approaches based on heavy quark symmetry and analytical constraints. Many
observables like Forward Backward (FB) asymmetry, single and double lepton
polarization asymmetries associated with the final state leptons, have been
extensively studied for quite some time for quark level processes $%
b\rightarrow s\left( d\right) \ell ^{+}\ell ^{-}$.

Recently, Belle\cite{15} has announced the first measurement of $%
B\rightarrow K_{1}^{+}(1270)\gamma $ 
\begin{equation}
\mathcal{B}(B^{+}\rightarrow K_{1}^{+}\gamma )=(4.28\pm 0.94\pm 0.43)\times
10^{-5}.  \label{01a}
\end{equation}
after which these radiative decays became topic of prime interest and their
is lot of theoretical progress for which one can see the Refs.\cite{16,17}.
In this paper we study the semileptonic $B$ meson decay $B\rightarrow
K_{1}\ell ^{+}\ell ^{-}$ using the framework of Gilani \textit{et al.}\cite
{18} with $K_{1}$ is an axial vector meson. The axial vector mesons is
distinguished by vector by an extra $\gamma _{5}$ in the gamma structure of
decay amplitude (DA) and some non perturbative parameters. But the presence
of extra $\gamma _{5}$ does not alter the calculation except the switching
of vector to axial vector form factors and vice a versa. As mentioned
earlier, the theoretical understanding of exclusive decays is complicated
mainly due to non-perturbative form factors entered in the long distance
non-perturbative contributions. The aim of this work is to relate the
various form factors in model independent way through Ward identities. This
enables us to make a clear separation between non-pole and pole type
contributions, the $q^{2}\rightarrow 0$ behavior of the former is known in
terms of a universal function $\xi _{\perp }\left( 0\right) \equiv
g_{+}\left( 0\right) $ introduced in the large energy effective theory
(LEET) of heavy ($B$) to light ($K_{1}$) form factors\cite{17}. The residue
of the pole is then determined in a self consistent way in terms of $%
g_{+}\left( 0\right) $ or $\xi _{\perp }\left( 0\right) $ which will give
information about the couplings of $B^{*}\left( 1^{-}\right) $ and $%
B_{A}^{*}\left( 1^{+}\right) $ with $BK_{1}$ channel. The from factors are
then determined in terms of the known parameters like $g_{+}\left( 0\right) $
and the masses of the particles involved which are then used to calculate
the branching ratio and forward-backward asymmetry for these decays.

This paper is organized as follows: In section II we introduce the effective
Hamiltonian formalism of semileptonic $B$ meson decays and will write down
the matrix elements for $B\rightarrow K_{1}\ell ^{+}\ell ^{-}$ decays.
Section III discusses the Ward identities and develop the relationship
between form factors which results in the reduction of number of unknown
quantities. The form factors thus obtained are used for the calculation of
decay width and forward-backward asymmetry. Finally, in the last section we
summarize our conclusions.

\section{Effective Hamiltonian}

At quark level the decay $B\rightarrow K_{1}\ell ^{+}\ell ^{-}$ is similar
to one studied in, for example, reference \cite{10}. The basic transition $%
b\rightarrow s\ell ^{+}\ell ^{-}$ is described by the effective Hamiltonian
given below 
\begin{equation}
H_{eff}=-4\frac{G_{F}}{\sqrt{2}}V_{tb}V_{ts}^{*}\sum_{i=1}^{10}C_{i}(\mu
)O_{i}(\mu )  \label{2.1}
\end{equation}
where $O_{i}$'$s$ are four local quark operators and $\dot{C}_{i}$ are
Wilson coefficients calculated in Naive dimensional regularization (NDR)
scheme \cite{20}.

One can write the above Hamiltonian in the following free quark decay
amplitude 
\begin{eqnarray}
\mathcal{M}(b &\rightarrow &s\ell ^{+}\ell ^{-})=\frac{G_{F}\alpha }{\sqrt{2}%
\pi }V_{tb}V_{ts}^{*}\left\{ 
\begin{array}{c}
C_{9}^{eff}\left[ \bar{s}\gamma _{\mu }Lb\right] \left[ \bar{\ell}\gamma
^{\mu }\ell \right] \\ 
+C_{10}\left[ \bar{s}\gamma _{\mu }Lb\right] \left[ \bar{\ell}\gamma ^{\mu
}\gamma ^{5}\ell \right] \\ 
-2\hat{m}_{b}C_{7}^{eff}\left[ \bar{s}i\sigma _{\mu \nu }\frac{\hat{q}^{\nu }%
}{\hat{s}}Rb\right] \left[ \bar{\ell}\gamma ^{\mu }\ell \right]
\end{array}
\right\}  \nonumber \\
&&  \label{2.2}
\end{eqnarray}
with $L/R\equiv \frac{\left( 1\mp \gamma _{5}\right) }{2}$, $s=q^{2}$ which
is just the momentum transfer from heavy to light meson. The amplitude given
in Eq.(\ref{2.2}) is a free quark decay amplitude which contains certain
long distance effect from the matrix element of local quark operators, $%
\left\langle l^{+}l^{-}s\left| O_{i}\right| b\right\rangle ,$ $1\leq i\leq
6, $ which usually reabsorbed into the redefinition of short distance Wilson
coefficients. Specifically, for the exclusive decays, the effective
coefficients of the operator $O_{9}=\frac{e^{2}}{16\pi ^{2}}(\bar{s}\gamma
_{\mu }Lb)(\bar{l}\gamma ^{\mu }l)$ can be written as 
\begin{equation}
C_{9}^{eff}=C_{9}+Y(\hat{s})  \label{2.3}
\end{equation}
where the perturbatively calculated result of $Y(\hat{s})$ is \cite{20,21} 
\begin{equation}
Y_{\text{pert}}\left( \hat{s}\right) =\left. 
\begin{array}{c}
g\left( \hat{m}_{c}\text{,}\hat{s}\right) \left(
3C_{1}+C_{2}+3C_{3}+C_{4}+3C_{5}+C_{6}\right) \\ 
-\frac{1}{2}g\left( 1\text{,}\hat{s}\right) \left(
4C_{3}+4C_{4}+3C_{5}+C_{6}\right) \\ 
-\frac{1}{2}g\left( 0\text{,}\hat{s}\right) \left( C_{3}+3C_{4}\right) +%
\frac{2}{9}\left( 3C_{3}+C_{4}+3C_{5}+C_{6}\right) .
\end{array}
\right.  \label{2.3a}
\end{equation}
For the values of the Wilson coefficients and the explicit expressions of $g$%
's appearing in Eq. (\ref{2.3a}) we will refer to \cite{20,21}. The hat
denotes normalization in terms of the $B$ meson mass\cite{10}.

\section{ Matix Elements and Ward Identities}

Exclusive decays $B\rightarrow K_{1}\ell ^{+}\ell $ involve the hadronic
matrix elements of the quark operators in Eq. (\ref{2.2}) between $B$ and $%
K_{1}.$ These can be parameterized in terms of form factors which are the
scalar functions of the four momentum square ($q^{2}=(p_{B}-p_{K_{1}})^{2}$%
). For the process we are considering, there are seven form factors like the
transition of psudoscalar to vector meson. The non-vanishing matrix elements
are 
\begin{eqnarray}
\left\langle K_{1}(k,\varepsilon )\left| V_{\mu }\right| B(p)\right\rangle
&=&i\varepsilon _{\mu }^{*}\left( M_{B}+M_{K_{1}}\right) V_{1}(s)  \nonumber
\\
&&-(p+k)_{\mu }\left( \varepsilon ^{*}\cdot q\right) \frac{V_{2}(s)}{%
M_{B}+M_{K_{1}}}  \nonumber \\
&&-q_{\mu }\left( \varepsilon \cdot q\right) \frac{2M_{K_{1}}}{s}\left[
V_{3}(s)-V_{0}(s)\right]  \label{3.1} \\
\left\langle K_{1}(k,\varepsilon )\left| A_{\mu }\right| B(p)\right\rangle
&=&\frac{2i\epsilon _{\mu \nu \alpha \beta }}{M_{B}+M_{K_{1}}}\varepsilon
^{*\nu }p^{\alpha }k^{\beta }A(s)  \label{3.2}
\end{eqnarray}
with $V_{\mu }=\bar{s}\gamma _{\mu }b$ and $A_{\mu }=\bar{s}\gamma _{\mu
}\gamma _{5}b$ are the vector and axial vector currents respectively and $%
\varepsilon _{\mu }^{*}$ is the polarization vector for the final state
axial vector meson. In Eq.(\ref{3.1})

\begin{equation}
V_{3}(s)=\frac{M_{B}+M_{K_{1}}}{2M_{K_{1}}}V_{1}(s)-\frac{M_{B}-M_{K_{1}}}{%
2M_{K_{1}}}V_{2}(s)  \label{3.2a}
\end{equation}
with 
\[
V_{3}(0)=V_{0}(0). 
\]
In addition to the above form factors there are also some penguin form
factors which are: 
\begin{eqnarray}
\left\langle K_{1}(k,\varepsilon )\left| \bar{s}i\sigma _{\mu \nu }q^{\nu
}b\right| B(p)\right\rangle &=&\left[ \left( M_{B}^{2}-M_{K_{1}}^{2}\right)
\varepsilon _{\mu }-(\varepsilon \cdot q)(p+k)_{\mu }\right] F_{2}(s) 
\nonumber \\
&&+(\varepsilon ^{*}\cdot q)\left[ q_{\mu }-\frac{s}{M_{B}^{2}-M_{K_{1}}^{2}}%
(p+k)_{\mu }\right] F_{3}(s)  \label{3.4} \\
\left\langle K_{1}(k,\varepsilon )\left| \bar{s}i\sigma _{\mu \nu }q^{\nu
}\gamma _{5}b\right| B(p)\right\rangle &=&-i\epsilon _{\mu \nu \alpha \beta
}\varepsilon ^{*\nu }p^{\alpha }k^{\beta }F_{1}(s)  \label{3.5}
\end{eqnarray}
with 
\[
F_{1}(0)=2F_{2}(0). 
\]
The various form factors appearing in Eqs. (\ref{3.1})-(\ref{3.5}) can be
related by Ward identities as follows \cite{18,23,24} 
\begin{eqnarray}
\left\langle K_{1}(k,\varepsilon )\left| \bar{s}i\sigma _{\mu \nu }q^{\nu
}b\right| B(p)\right\rangle &=&-(m_{b}+m_{s})\left\langle
K_{1}(k,\varepsilon )\left| \bar{s}\gamma _{\mu }b\right| B(p)\right\rangle
\label{3.6} \\
\left\langle K_{1}(k,\varepsilon )\left| \bar{s}i\sigma _{\mu \nu }q^{\nu
}\gamma _{5}b\right| B(p)\right\rangle &=&(m_{b}-m_{s})\left\langle
K_{1}(k,\varepsilon )\left| \bar{s}\gamma _{\mu }\gamma _{5}b\right|
B(p)\right\rangle  \nonumber \\
&&+(p+k)_{\mu }\left\langle K_{1}(k,\varepsilon )\left| \bar{s}\gamma
_{5}b\right| B(p)\right\rangle .  \label{3.8}
\end{eqnarray}
Now we make the heavy quark approximation and compare coefficients of $%
\varepsilon _{\mu }^{*}$ and $q_{\mu }$ from both sides. In the heavy quark
approximation we need not to compare the coefficients $(p+k)_{\mu }.$ Using
Eqs. (\ref{3.1})-(\ref{3.5}) in Eqs. (\ref{3.6}) and (\ref{3.8}), we get the
following relationship between form factors 
\begin{eqnarray}
F_{1}(s) &=&\ -\frac{(m_{b}-m_{s})}{M_{B}+M_{K_{1}}}2A(s)  \label{3.9} \\
F_{2}(s) &=&-\frac{(m_{b}+m_{s})}{M_{B}-M_{K_{1}}}V_{1}(s)  \label{3.10} \\
F_{3}(s) &=&\frac{2M_{K_{1}}}{s}(m_{b}+m_{s})\left[ V_{3}(s)-V_{0}(s)\right]
.  \label{3.11}
\end{eqnarray}
These are model independent results derived by using Ward identities. The
universal normalization of the above form factors at $q^{2}=s=0$ is obtained
by defining\cite{18} 
\begin{eqnarray}
\left\langle K_{1}(k,\varepsilon )\left| \bar{s}i\sigma ^{\alpha \beta
}\gamma ^{5}b\right| B(p)\right\rangle &=&-i\epsilon ^{\alpha \beta \rho
\sigma }\varepsilon _{\rho }^{*}\left[ (p+k)_{\rho }g_{+}+q_{\sigma
}g_{-}\right] -(q\cdot \varepsilon ^{*})\epsilon ^{\alpha \beta \rho \sigma
}(p+k)_{\rho }q_{\sigma }h  \nonumber \\
&&-i\left[ q^{\alpha }\epsilon ^{\beta \rho \sigma \tau }\varepsilon _{\rho
}^{*}(p+k)_{\sigma }q_{\tau }-\alpha \leftrightarrow \beta \right] h_{1}.
\label{3.12}
\end{eqnarray}

Using Dirac identity 
\begin{equation}
\sigma _{\mu \nu }\gamma ^{5}=-\frac{i}{2}\epsilon _{\mu \nu \alpha \beta
}\sigma ^{\alpha \beta }  \label{3.13}
\end{equation}
in Eq. (\ref{3.12}) one can write 
\begin{eqnarray}
\left\langle K_{1}(k,\varepsilon )\left| \bar{s}i\sigma _{\mu \nu }q^{\nu
}b\right| B(p)\right\rangle &=&\varepsilon _{\mu }^{*}\left[ \left(
M_{B}^{2}-M_{K_{1}}^{2}\right) g_{+}+sg_{-}\right] -(q\cdot \varepsilon
^{*})\left[ (p+k)_{\mu }g_{+}+q_{\mu }g_{-}\right]  \nonumber \\
&&+(q\cdot \varepsilon ^{*})\left[ s(p+k)_{\mu
}-(M_{B}^{2}-M_{K_{1}}^{2})q_{\mu }\right] h.  \label{3.13a}
\end{eqnarray}
Comparing coefficients of $q_{\mu }$, $\varepsilon _{\mu }^{*}$ and $%
\epsilon _{\mu \nu \alpha \beta }$ from Eqs.(\ref{3.4}), (\ref{3.5}) and
Eqs. (\ref{3.12}) and (\ref{3.13a}), we get 
\begin{eqnarray}
F_{1}(s) &=&2\left[ g_{+}(s)-sh_{1}\right]  \label{3.14} \\
F_{2}(s) &=&g_{+}+\frac{s}{M_{B}^{2}-M_{K_{1}}^{2}}g_{-}  \label{3.15} \\
F_{3}(s) &=&-g_{-}-\left( M_{B}^{2}-M_{K_{1}}^{2}\right) h.  \label{3.16}
\end{eqnarray}
The above results ensure that $F_{1}(0)=2F_{2}(0)$. In terms of $g_{+}$, $%
g_{-}$ and $h$, the form factors become 
\begin{eqnarray}
A(s) &=&\frac{M_{B}+M_{K_{1}}}{m_{b}-m_{s}}\left[ g_{+}(s)-sh_{1}\right] 
\nonumber \\
V_{1}(s) &=&-\frac{M_{B}-M_{K_{1}}}{m_{b}+m_{s}}\left[ g_{+}+\frac{s}{%
M_{B}^{2}-M_{K_{1}}^{2}}g_{-}\right]  \nonumber \\
V_{2}(s) &=&-\left( \frac{M_{B}+M_{K_{1}}}{m_{b}+m_{s}}\right) \left[
g_{+}(s)-sh\right] -\frac{2M_{K_{1}}}{M_{B}-M_{K_{1}}}V_{0}(s).
\label{3.16a}
\end{eqnarray}
By looking at the above expressions one can see that the normalization of
above form factors $A$ and $V_{1}$ at $s=0$ is determined by the single
constant $g_{+}\left( 0\right) $ where as that of $V_{2}$ is determined by $%
g_{+}\left( 0\right) $ and $V_{0}(s)$.

\subsection{Pole contributions}

The pole contribution for $B$ to $\rho $ has been studied in detail by
Gilani \textit{et al}.\cite{18}. This remains the same for $B$ to $K_{1}$
transition and again only $h_{1}$, $g_{-}$, $h$ and $V_{0}$ get pole
contributions from $B^{*}(1^{-}),B_{A}^{*}(1^{+})$ and $B(0^{-})$ mesons
where as $g_{+}$, $g_{-}$ and $V_{0}(s)$ gets their contribution from quark
triangle graph. These are given by 
\begin{eqnarray}
h_{1}|_{pole} &=&-\frac{1}{2}\frac{g_{B^{*}BK_{1}}}{M_{B^{*}}^{2}}\frac{%
f_{T}^{B^{*}}}{1-s/M_{B^{*}}^{2}}=\frac{R_{V}}{M_{B^{*}}^{2}}\frac{1}{%
1-s/M_{B^{*}}^{2}}  \nonumber \\
g_{-}|_{pole} &=&-\frac{g_{B_{A}^{*}BK_{1}}}{M_{B_{A}^{*}}^{2}}\frac{%
f^{B_{A}^{*}}}{1-s/M_{B_{A}^{*}}^{2}}=\frac{R_{A}^{S}}{M_{B_{A}^{*}}^{2}}%
\frac{1}{1-s/M_{B_{A}^{*}}^{2}}  \nonumber \\
h|_{pole} &=&\frac{1}{2}\frac{f_{B_{A}^{*}BK_{1}}}{M_{B_{A}^{*}}^{2}}\frac{%
f_{T}^{B_{A}^{*}}}{1-s/M_{B_{A\;}^{*}}^{2}}=\frac{R_{A}^{D}}{%
M_{B_{A}^{*}}^{2}}\frac{1}{1-s/M_{B_{A}^{*}}^{2}}  \nonumber \\
&&V_{0}(s)|_{pole}=\frac{g_{BBK_{1}}}{M_{K_{1}}}f_{B}\frac{s/M_{B}^{2}}{%
1-s/M_{B}^{2}}=R_{0}\frac{s/M_{B}^{2}}{1-s/M_{B}^{2}}  \label{pole1}
\end{eqnarray}
where $R_{V}$, $R_{A}^{S}$, $R_{A}^{D}$ and $R_{0}$ are related to the
coupling constants $g_{B^{*}BK_{1}}$, $g_{B_{A}^{*}BK_{1}}$, $%
f_{B_{A}^{*}BK_{1}}$ and $g_{BBK_{1}}$ respectively. One can find the detail
about it in Ref.\cite{18}. Thus one can write 
\begin{eqnarray}
A(s) &=&\left( \frac{M_{B}+M_{K_{1}}}{m_{b}-m_{s}}\right) \left(
g_{+}(s)-R_{V}\frac{s}{M_{B^{*}}^{2}}\left( \frac{1}{1-s/M_{B^{*}}^{2}}%
\right) \right)  \label{3.17} \\
V_{1}(s) &=&-\left( \frac{M_{B}-M_{K_{1}}}{m_{b}+m_{s}}\right) \left(
g_{+}(s)+\frac{s}{M_{B}^{2}-M_{K_{1}}^{2}}\tilde{g}_{-}+\frac{R_{A}^{S}}{%
M_{B}^{2}-M_{K_{1}}^{2}}\frac{s}{M_{B_{A}^{*}}^{2}}\left( \frac{1}{%
1-s/M_{B_{A}^{*}}^{2}}\right) \right)  \nonumber \\
&&  \label{3.18} \\
V_{2}(s) &=&-\left( \frac{M_{B}+M_{K_{1}}}{m_{b}+m_{s}}\right) \left[
g_{+}(s)-\frac{s}{M_{B_{A}^{*}}^{2}}R_{A}^{D}\frac{1}{1-s/M_{B_{A}^{*}}^{2}}%
\right] -\frac{2M_{K_{1}}}{M_{B}-M_{K_{1}}}V_{0}(s).  \label{3.19}
\end{eqnarray}
The behavior of $g_{+}(s),\tilde{g}_{-}(s)$ and $V_{0}(s)$ near $%
s\rightarrow 0$ is known from LEET and their form is\cite{18} 
\begin{eqnarray}
g_{+}(s) &=&\frac{\xi _{\perp }(0)}{\left( 1-s/M_{B}^{2}\right) ^{2}}=-%
\tilde{g}_{-}(s)  \label{3.20} \\
V_{0}(s) &=&\left( 1-\frac{M_{K_{1}}^{2}}{M_{B}E_{K_{1}}}\right) \xi _{\Vert
}(s)+\frac{M_{K_{1}}}{M_{B}}\xi _{\bot }(s).  \label{3.21}
\end{eqnarray}
At $s\rightarrow 0$%
\begin{eqnarray}
V_{0}(0) &=&\frac{M_{B}^{2}-M_{K_{1}}^{2}}{M_{B}^{2}+M_{K_{1}}^{2}}\xi
_{\Vert }(0)+\frac{M_{K_{1}}}{M_{B}}\xi _{\bot }(0)  \label{3.22} \\
E_{K_{1}} &=&\frac{M_{B}}{2}\left( 1-\frac{s}{M_{B}^{2}}+\frac{M_{K_{1}}^{2}%
}{M_{B}^{2}}\right)  \label{3.22a} \\
g_{+}(0) &=&\xi _{\perp }(0).  \label{3.23}
\end{eqnarray}
The pole terms in the relations (\ref{3.17}), (\ref{3.18}) and (\ref{3.19})
are expected to dominate near $s=M_{B^{*}}^{2}$ or $M_{B_{A}^{*}}^{2}$. On
the other hand the relations obtained from Ward identities, are expected to
hold for $s$ much below the resonance region. The above behavior, near $s=0$
and that near the pole\cite{18} suggest 
\begin{equation}
F(s)=\frac{F(0)}{\left( 1-s/M^{2}\right) (1-s/M^{\prime 2})}  \label{3.24}
\end{equation}
where $M^{2}$ is $M_{B^{*}}^{2}\,$or $M_{B_{A}^{*}}^{2}$ and $M^{\prime }$
is the radial excitation of $M$. This parameterization not only takes into
account the corrections to the single pole dominance, as suggested by
dispersion relation \cite{23,24,25}, but also of off-mass-shell-ness of
couplings of $B^{*}$ or $B_{A}^{*}$ with $BK_{1}$ channel.

Since $g_{+}(s)$ and $\tilde{g}_{-}(s)$ have no pole at $s=M_{B^{*}}^{2}$,
therefore we get 
\[
A(s)\left( 1-\frac{s}{M_{B^{*}}^{2}}\right) |_{s=M_{B^{*}}^{2}}=R_{V}\left( 
\frac{M_{B}+M_{K_{1}}}{m_{b}-m_{s}}\right) 
\]
This gives using the parametrization (\ref{3.24}) 
\begin{equation}
R_{V}\equiv -\frac{1}{2}g_{B^{*}BK_{1}}f_{T}^{B^{*}}=-\frac{1}{2}%
g_{B^{*}BK_{1}}f_{B}=-\frac{g_{+}(0)}{1-M_{B^{*}}^{2}/M_{B^{*}}^{\prime 2}}
\label{3.25}
\end{equation}
Similarly, 
\begin{equation}
R_{A}^{D}\equiv \frac{1}{2}f_{B_{A}^{*}BK_{1}}f_{T}^{B_{A}^{*}}=-\frac{%
g_{+}(0)}{1-M_{B_{A}^{*}}^{2}/M_{B_{A}^{*}}^{\prime 2}}.  \label{3.26}
\end{equation}
For the detailed derivation and discussion on these relations we will refer
to \cite{18}. We cannot use the parametrization given in Eq.(\ref{3.24}) for 
$V_{1}(s)$ since near $s=0,$ $V_{1}(s)$ behaves as $g_{+}(s)\left[
1-s/(M_{B}^{2}-M_{K_{1}}^{2})\right] $ [c.f. Eqs. (\ref{3.18}) and (\ref
{3.20})]. This suggests the following 
\begin{equation}
V_{1}(s)=\frac{g_{+}(0)}{\left( 1-s/M_{B_{A}^{*}}^{2}\right) \left(
1-s/M_{B_{A}^{*}}^{\prime 2}\right) }\left( 1-\frac{s}{%
M_{B}^{2}-M_{K_{1}}^{2}}\right) .  \label{3.27}
\end{equation}
Until now we have expressed every thing in terms of $g_{+}(0)$ which is the
only unknown in the calculation. After the first announcement of Belle\cite
{15} for the decay $B\rightarrow K_{1}\gamma $, the value of $g_{+}(0)$ has
been extracted to be \cite{16,17} 
\begin{equation}
g_{+}(0)=\xi _{\perp }(0)=0.32\pm 0.03.  \label{3.28}
\end{equation}
Using $f_{B}=180$ MeV we have the prediction from Eq. (\ref{3.25}) 
\begin{equation}
g_{B^{*}BK_{1}}=15.42\text{ GeV}^{-1}  \label{3.28a}
\end{equation}
Similarly, the $S$ and $D$ wave couplings are predicted to be 
\begin{equation}
g_{B_{A}^{*}BK_{1}}=3.17f_{B_{A}^{*}BK_{1}}\text{GeV}^{2}  \label{3.28b}
\end{equation}
The different values of $F(0)$'s are 
\begin{eqnarray}
A(0) &=&\left( \frac{M_{B}+M_{K_{1}}}{m_{b}-m_{s}}\right) g_{+}(0)
\label{3.29} \\
V_{1}(0) &=&-\left( \frac{M_{B}-M_{K_{1}}}{m_{b}+m_{s}}\right) g_{+}(0)
\label{3.30} \\
V_{2}(0) &=&-\left( \frac{M_{B}+M_{K_{1}}}{m_{b}+m_{s}}\right) g_{+}(0)-%
\frac{2M_{K_{1}}}{M_{B}-M_{K_{1}}}V_{0}(0)  \label{3.31}
\end{eqnarray}
where $g_{+}(0)$ is the same as given in Eq. (\ref{3.28}). The calculation
of numerical values of $A(0),V_{1}(0)$ is very trivial but to go for $%
V_{2}(0)$, we have to know the value of $V_{0}\left( 0\right) $. Although
LEET does not give any relationship between $\xi _{\parallel }(0)$ and $\xi
_{\perp }(0)$ but due to some numerical coincidence in the LCSR expressions
for $\xi _{\parallel }(0)$ and $\xi _{\perp }(0)$\cite{26} 
\begin{equation}
\xi _{\parallel }(0)\simeq \xi _{\perp }(0)=g_{+}(0)  \label{3.32}
\end{equation}
so that, from Eq. (\ref{3.22}) 
\begin{equation}
V_{0}(0)=1.13g_{+}(0).  \label{3.33}
\end{equation}
Thus the final expressions of form factors which we shall use for numerical
work are 
\begin{eqnarray}
A\left( s\right) &=&\frac{A\left( 0\right) }{\left( 1-s/M_{B}^{2}\right)
(1-s/M_{B}^{\prime 2})}  \nonumber \\
V_{1}(s) &=&-\frac{V_{1}(0)}{\left( 1-s/M_{B_{A}^{*}}^{2}\right) \left(
1-s/M_{B_{A}^{*}}^{\prime 2}\right) }\left( 1-\frac{s}{%
M_{B}^{2}-M_{K_{1}}^{2}}\right)  \label{3.34} \\
V_{2}(s) &=&-\frac{\tilde{V}_{2}(0)}{\left( 1-s/M_{B_{A}^{*}}^{2}\right)
\left( 1-s/M_{B_{A}^{*}}^{\prime 2}\right) }-\frac{2M_{K_{1}}}{%
M_{B}-M_{K_{1}}}\frac{V_{0}(0)}{\left( 1-s/M_{B}^{2}\right) \left(
1-s/M_{B}^{\prime 2}\right) }  \nonumber
\end{eqnarray}
where 
\begin{eqnarray}
A(0) &=&(0.52\pm 0.05)  \nonumber \\
V_{1}(0) &=&-(0.24\pm 0.02)  \nonumber \\
\tilde{V}_{2}(0) &=&-(0.39\pm 0.03)  \label{3.35}
\end{eqnarray}

\section{Decay Distribution and Forward-Backward Asymmetry}

In this section we define the decay rate distribution which we shall use for
the phenomenological analysis . Following the notation from ref.\cite{10} we
can write from Eq. (\ref{2.2}) 
\begin{equation}
\mathcal{M}=\frac{G_{F}\alpha }{2\sqrt{2}\pi }V_{tb}V_{ts}^{*}m_{B}\left[ 
\mathcal{T}_{\mu }^{1}\left( \bar{l}\gamma ^{\mu }l\right) +\mathcal{T}_{\mu
}^{2}\left( \bar{l}\gamma ^{\mu }\gamma ^{5}l\right) \right]  \label{4.1}
\end{equation}
where 
\begin{eqnarray}
\mathcal{T}_{\mu }^{1} &=&A\left( \hat{s}\right) \varepsilon _{\mu \rho
\alpha \beta }\epsilon ^{*\rho }\hat{p}_{B}^{\alpha }\hat{p}_{K_{1}}^{\beta
}-iB\left( \hat{s}\right) \epsilon _{\mu }^{*}+iC\left( \hat{s}\right)
\left( \epsilon ^{*}\cdot \hat{p}_{B}\right) \hat{p}_{h\mu }+iD\left( \hat{s}%
\right) \left( \epsilon ^{*}\cdot \hat{p}_{B}\right) \hat{q}_{\mu } 
\nonumber \\
&&  \label{4.2} \\
\mathcal{T}_{\mu }^{2} &=&E\left( \hat{s}\right) \varepsilon _{\mu \rho
\alpha \beta }\epsilon ^{*\rho }\hat{p}_{B}^{\alpha }\hat{p}_{K_{1}}^{\beta
}-iF\left( \hat{s}\right) \epsilon _{\mu }^{*}+iG\left( \hat{s}\right)
\left( \epsilon ^{*}\cdot \hat{p}_{B}\right) \hat{p}_{h\mu }+iH\left( \hat{s}%
\right) \left( \epsilon ^{*}\cdot \hat{p}_{B}\right) \hat{q}_{\mu } 
\nonumber \\
&&  \label{4.3}
\end{eqnarray}
The definition of different momenta involved are defined in reference\cite
{10}, where the auxiliary functions are 
\begin{eqnarray}
A(\hat{s}) &=&-\frac{2A(\hat{s})}{1+\hat{M}_{K_{1}}}C_{9}^{eff}(\hat{s})+%
\frac{2\hat{m}_{b}}{\hat{s}}C_{7}^{eff}F_{1}(\hat{s})  \nonumber \\
B(\hat{s}) &=&\left( 1+\hat{M}_{K_{1}}\right) \left[ C_{9}^{eff}(\hat{s}%
)V_{1}(\hat{s})+\frac{2\hat{m}_{b}}{\hat{s}}C_{7}^{eff}\left( 1-\hat{M}%
_{K_{1}}\right) \right]  \nonumber \\
C\left( \hat{s}\right) &=&\frac{1}{\left( 1-\hat{M}_{K_{1}}^{2}\right) }%
\left\{ C_{9}^{eff}(\hat{s})V_{2}(\hat{s})+2\hat{m}_{b}C_{7}^{eff}\left[
F_{3}(\hat{s})+\frac{1-\hat{M}_{K_{1}}^{2}}{\hat{s}}F_{2}(\hat{s})\right]
\right\}  \nonumber \\
D(\hat{s}) &=&\frac{1}{\hat{s}}\left[ 
\begin{array}{c}
\left( C_{9}^{eff}(\hat{s})(1+\hat{M}_{K_{1}})V_{1}(\hat{s})-(1-\hat{M}%
_{K_{1}})V_{2}(\hat{s})-2\hat{M}_{K_{1}}V_{0}(\hat{s})\right) \\ 
-2\hat{m}_{b}C_{7}^{eff}F_{3}(\hat{s})
\end{array}
\right]  \nonumber \\
E(\hat{s}) &=&-\frac{2A(\hat{s})}{1+\hat{M}_{K_{1}}}C_{10}  \nonumber \\
F(\hat{s}) &=&\left( 1+\hat{M}_{K_{1}}\right) C_{10}V_{1}(\hat{s})  \nonumber
\\
G(\hat{s}) &=&\frac{1}{1+\hat{M}_{K_{1}}}C_{10}V_{2}(\hat{s})  \nonumber \\
H(\hat{s}) &=&\frac{1}{\hat{s}}\left[ C_{10}(\hat{s})(1+\hat{M}%
_{K_{1}})V_{1}(\hat{s})-(1-\hat{M}_{K_{1}})V_{2}(\hat{s})-2\hat{M}%
_{K_{1}}V_{0}(\hat{s})\right] .  \label{4.4}
\end{eqnarray}
The differential decay rate for $B\rightarrow K^{*}\mu ^{+}\mu ^{-}$ can be
expressed in terms of these auxiliary functions in\cite{10} and this remains
the same for $B\rightarrow K_{1}\mu ^{+}\mu ^{-}$ with the obvious
replacements. Integration on $\hat{s}$ in the range 
\begin{equation}
\left( 2\hat{m}_{l}\right) ^{2}\leq \hat{s}\leq \left( 1-\hat{m}%
_{K_{1}}\right) ^{2}  \label{4.5}
\end{equation}
with $\hat{m}_{l}=m_{l}/m_{B}$, and using $\tau _{B^{0}}=\left( 1.530\pm
0.009\right) \times 10^{-12}s$, the branching ratio is 
\[
\mathcal{B}\left( B\rightarrow K_{1}\mu ^{+}\mu ^{-}\right)
=0.9_{-\,0.14}^{+\,0.11}\times 10^{-7} 
\]
\newline
The above value of branching ratio is for the case if we do not include $Y(%
\hat{s})$ in Eq. (\ref{2.3}). The error in the value reflects the
uncertainty from the form factors, and due to the variation of input
parameters like CKM matrix elements, decay constant of $B$ meson and masses
as defined in Table I.

\begin{eqnarray*}
\text{Table I} &:&\text{Default value of input parameters used in the
calculation} \\
&& \\
&& 
\begin{tabular}{ll}
\hline
$m_{W}$ & $80.41$ GeV \\ 
$m_{Z}$ & $91.1867$ GeV \\ 
$sin^{2}\theta _{W}$ & $0.2233$ \\ 
$m_{c}$ & $1.4$ GeV \\ 
$m_{b,pole}$ & $4.8\pm 0.2$ GeV \\ 
$m_{t}$ & $173.8\pm 5.0$ GeV \\ 
$\alpha _{s}\left( m_{Z}\right) $ & $0.119\pm 0.0058$ \\ 
$f_{B}$ & $\left( 200\pm 30\right) $ MeV \\ 
$\left| V_{ts}^{*}V_{tb}\right| $ & $0.0385$ \\ \hline
\end{tabular}
\end{eqnarray*}

Now if we include the value of $Y(\hat{s})$ the central value of branching
ratio reduces to 
\[
\mathcal{B}\left( B\rightarrow K_{1}\mu ^{+}\mu ^{-}\right) =0.72\times
10^{-7} 
\]
By including $Y(\hat{s})$ the behavior of the differential decay rate as a
function of $\hat{s}$ is shown in Fig. 1$.$ The solid line denotes the
theoretical prediction with input parameters taken at their central values,
while the band between two dashed line shows the uncertainity from input
parameters. In our numerical analysis we have considered only the final
state leptons as being the muon. Our reason for choosing this is due to the
extreme difficulty in detecting electron in the final state and that the
branching ratio $B\rightarrow K_{1}\ell ^{+}\ell ^{-}$ becoming small with
the SM for the $\tau $ in the final state.

The differential forward-backward asymmetry for $B\rightarrow K_{1}\mu
^{+}\mu ^{-}$ reads as follows\cite{10} 
\begin{equation}
\frac{d\mathcal{A}_{\text{FB}}}{d\hat{s}}=\frac{G_{F}^{2}\alpha ^{2}m_{B}^{5}%
}{2^{10}\pi ^{5}}\left| V_{ts}^{*}V_{tb}\right| ^{2}\hat{s}\hat{u}\left( 
\hat{s}\right) \left[ \text{Re}\left( BE^{*}\right) +\text{Re}\left(
AF^{*}\right) \right]  \label{4.6}
\end{equation}
where 
\begin{eqnarray}
\hat{u}\left( \hat{s}\right) &=&\sqrt{\lambda \left( 1-4\frac{\hat{m}_{l}^{2}%
}{\hat{s}}\right) }  \nonumber \\
\lambda &\equiv &\lambda \left( 1,\hat{m}_{K_{1}}^{2},\hat{s}\right) 
\nonumber \\
&=&1+\hat{m}_{K_{1}}^{4}+\hat{s}^{2}-2\hat{s}-2\hat{m}_{K_{1}}^{2}\left( 1+%
\hat{s}\right)  \label{4.7}
\end{eqnarray}
The variable $\hat{u}$ corresponds to $\theta $, the angle between the
momentum of the $B$ meson and the positively charged lepton in the dilepton
c.m. system frame. The behavior of forward-backward asymmetry in $%
B\rightarrow K_{1}\mu ^{+}\mu ^{-}$ decay as a function of $\hat{s}$ is
shown in Fig. 2. Contrary to the branching ratio, the forward-backward
asymmetry is less sensitive to the input parameters as is clear from Fig. 2.
For the zero-point of forward-backward asymmetry in the standard model, we
get $\hat{s}=\left( 0.16+0.01\right) $ $\left( s=\left( 4.46+0.27\right) 
\text{ GeV}^{-2}\right) $.

\textbf{Conclusions}

We have studied $B\rightarrow K_{1}\ell ^{+}\ell ^{-}$ decay using Ward
identities. The form factors have been calculated and found that their
normalization is essentially determined by single constant $g_{+}\left(
0\right) $ which has the value $g_{+}(0)=0.32\pm 0.03$ obtained from \cite
{16,17}. By considering the radial excitation of $M$ (where $M=M_{B^{*}\text{
}}$or $M_{B_{A}^{*}\text{ }}$), which are suggested by dispersion relation%
\cite{18}, we have predicted the coupling of $B^{*}$ or $B_{A}^{*}$ with $%
BK_{1}$ channel as indicated in Eq. (\ref{3.28a}) and the value is $%
g_{B^{*}BK_{1}}=15.42$ GeV$^{-1}$. Also we have predicted the relationship
between $S$ and $D$ wave couplings $%
g_{B_{A}^{*}BK_{1}}=3.17f_{B_{A}^{*}BK_{1}}$GeV$^{2}$ given in Eq. (\ref
{3.28b}). We have summarized our form factors in Eq. (\ref{3.34}) and their
value at $s=0$ in Eq. (\ref{3.35}). By using these form factors we have
calculated the branching ratio for $B\rightarrow K_{1}\mu ^{+}\mu ^{-}$ both
by considering the non resonant and resonant value of the Wilson coefficient 
$C_{9}^{eff}(\hat{s})$ which will been seen in future experiments. The decay
distribution is shown graphically in\ Fig. 1, where the differential decay
rate is plotted as a function $\hat{s}$.

A detailed analysis of the forward-backward asymmetry is also presented
here. We have plotted the forward-backward asymmetry as a function of $\hat{s%
}$ in Fig. 2. It is clear from the graph that the the SM the central value
of the zero of the FB asymmetry is at $\hat{s}=0.16$ $\left( s=4.46\right) $%
. This value of the zero of the forward-backward asymmetry will provide the
precision test of SM in planned future experiments.

\textbf{Acknowledgments}

The authors would like to thank Prof.\ Riazuddin and Prof. Fayyazuddin for
useful discussion. The work of Ali and Ishtiaq was supported by the World
Lab. fellowship.

\textbf{Figure Captions}

1): The differential decay rate as a function of $\hat{s}$ is plotted using
the form factors calculated by using Ward Identities. The resonanant $c\bar{c%
}$ states are parameterized as in refs.\cite{20,21}. Here the solid line
denotes the theoretical predictions with the input parameters taken at their
central values, while the dashed (dotted) line is for max. (min) value of
input parameters.

2): The forward-backward (FB) asymmetry as a function of $\hat{s}$ is
plotted using the form factors calculated by using Ward Identities. The
resonanant $c\bar{c}$ states are parameterized as in refs.\cite{20,21}. The
dashed (solid) line is for the central (max.) value of the input paramteres.

\end{document}